\documentclass[twoside,12pt]{article}
\usepackage{amssymb,latexsym}
\usepackage{amsfonts,euscript}
\begin{document}
\title{\huge{\bf{Singletons and Neutrinos }}}
\author{{\sc{Christian Fr\o nsdal}}\\
{\it Physics Department, University of California,}\\
{\it Los Angeles, CA 90095-1547 USA}\\
{\small{e-mail:}} {\tt {\small{fronsdal@physics.ucla.edu}}}}
\date{}
\maketitle
\vskip-5mm
\centerline{\it Dedicated to the memory of my friend Mosh\'e Flato.}
\thispagestyle{empty}
 \begin{abstract}
The first half is a rapid review of 30 years of work on
physics in anti-De Sitter  space, with heavy emphasis on singletons.
Principal topics are the kinematical basis for
regarding singletons as the constituents of massless particles, and the
effect of (negative) curvature in the infrared domain. Ideas that lead
to an alternative to Big Bang cosmology are merely sketched.
The second half presents new ideas inspired by experimental results on
neutrino oscillations. Since leptons are massless before symmetry breaking
it is natural to view them as composite states consisting of one Bose
singleton (the Rac) and one Fermi singleton (the Di). This gives rise
to a particular formulation of the phenomenology of electroweak physics,
and strong suggestions for an expansion of the Standard model. An expansion
of the Higgs sector seems inevitable, and flavor changing symmetry,
complete with a new set of heavy vector mesons, is a very attractive
possibility.
\end{abstract}
%%%%%% again kluwer style %%%%
%%\dedication{Dedicated to the memory of my friend Mosh\'e Flato.}
%%\begin{keywords} keywords
%%\end{keywords}
%%\classification{Mathematics Subject Classifications (1991)}
%%\end{opening}
%%%%%%%%%%%%%%%%%%%%%%%%%%%%%%%%
\section{Introduction}

This talk is intended as a modest offering to the memory of Mosh\'e Flato,
the most original physicist that I have known, and my best friend for more
than 25 years.

Perhaps you will say that Mosh\'e was a mathematician as much as he was a
physicist.  Indeed, he was a professor of mathematics and he created, from
nothing, one of the finest departments of mathematical physics anywhere;
nevertheless,   his great passion was physics.  Others will tell you
about some of Mosh\'e's greatest accomplishments in mathematics and in
mathematical physics; it falls on me to summarize his work on the
{\bf physics} of spaces of constant curvature, an endeavor that is
entirely in the realm of physics.

As it happens, this work, on which I shall report, was done in
collaboration with others, including Daniel Sternheimer and myself, and the
early part of it was initiated before we started to publish together,
but Mosh\'e's influence dates back to the beginning.

\section{Physics in Anti-De Sitter Space}

The initial problematics was very simple: The cosmological constant is small,
but it can never be shown, experimentally, to be exactly zero.  Any value
other than zero is incompatible with the idea that the Poincar\'e group
is the basic symmetry group of space time, but it is consistent with a symmetry
group that has the same dimension as the Poincar\'e group;
all of the most important concepts of flat space physics can be extended to
a space of constant curvature that admits the global action of one of the
two De Sitter groups, $SO(3,2)$ or $SO(4,1)$.

This idea was not new in the mid-sixties (see e.g. \cite{W50,GL,N67}).
But up to that time, of the two possibilities, $SO(4,1)$ had received the
most attention although it was known, already, that symmetry under this
group leads unavoidably to spontaneous creation of matter (see e.g.
\cite{N68}).

Both of us had a great admiration for Wigner's work \cite{W39} on the unitary
representations of the Poincar\'e group that led to the first classification
of elementary particles, in terms of mass and spin (or helicity). It was
very natural to apply the same philosophy to the classification of elementary
particles in (anti-) De Sitter space.  The unitary representations of both
groups were essentially known at the time (cf. \cite{T,Dx} for $SO(4,1)$
and \cite{E} for the discrete series of $SO(3,2)$; the full unitary
dual was calculated only at the end of the seventies \cite{A79}), but the
interpretation in terms of elementary particles was not yet well developed.

When the problem is approached in this way, then immediately one is led to
favor $SO(3,2)$, the symmetry group of anti-De Sitter space (so called later
on, and now abbreviated as AdS), over the other alternative.  This is
because $SO(3,2)$ has representations that can be associated with
elementary particles, while $SO(4,1)$ does not.  The energy spectrum of
every unitary representation of $SO(4,1)$, except the trivial one, is the
real line, unbounded in both directions \cite{S68}, and if there is one
physical principle that has stood the test of time, it is the requirement
that energy must be bounded below. Consequently, all our work was
concerned exclusively with $SO(3,2)$ and anti-De Sitter space time.

The choice of $SO(3,2)$ implies a negative curvature. This curvature, in the
cosmological context, is very small, and it was never expected to be
measurable.  Our project was not concerned with the \emph{magnitude} of the
cosmological constant, but merely with the fact that, as a matter of
principle, it \emph{need not be zero}.  It was expected that physics in a
space of constant, negative curvature was possible and above all that its
elaboration would be very educational.  That it would actually turn out
to suggest new types of physical phenomena was a great surprise.

The first indication of new physics appeared in connection with ``massless''
particles.  It turns out that ``masslessness'' is a term that can be applied
to certain irreducible representations of $SO(3,2)$ with as much justification
as in the context of the Poincar\'e group \cite{AFFS}.  One instance of
masslessness appears in anti-De Sitter electrodynamics \cite{F75}.

For the sake of simplicity, I shall not deal directly with realistic
Maxwell theory, but instead with the theory of a spinless, massless field.
All the interesting features of a gauge theory are then lost, but the
features that I want to discuss are not. (The realistic picture involves
Gupta-Bleuler triplets and indecomposable representations \cite{F75,FF88}.)
In flat space free (spinless) photons are associated with an irreducible
representation  of the Poincar\'e group; this particular representation
has the interesting property of having a unique extension to the conformal
group \cite{AF78}.  The conformal group in anti-De Sitter space time is
the same as the conformal group in flat space, locally isomorphic to
$SO(4,2)$, and the same irreducible representation of the conformal group
appears in (scalar) electrodynamics, in both cases.
But in the case of anti-De Sitter space this irreducible representation
breaks up into two inequivalent representations of $SO(3,2)$.  This means
that there are two kinds of photons, with different propagators.
In order to incorporate conformal invariance into the theory of free
quantum fields, one must quantize the field in such a way that both
types propagate.  The big surprise is that this is incompatible with
the self-adjointness of the Hamiltonian.  In other words, energy is not
conserved.  To conserve energy, one must use only one of the two types
of photons; this amounts to spontaneous breakdown of conformal symmetry.

Now I should like to present here for the first time an idea that would have
been elaborated by Mosh\'e and myself if Mosh\'e had been given more time.
It is a radical idea, for it suggests an alternative to the Big Bang, and
thus it is likely to irritate some people; but I will risk it.  The idea is
to accept the lack of energy conservation that is implied by conformal
invariance of QED. This will lead to spontaneous creation of energy, locally
throughout the universe. Conceivably, the amount of energy (and matter)
created could balance the loss occasioned by the divergence of matter in
the form of visible galaxies, and lead to a kind of ``steady state'' model of
cosmology, characterized by a mean mass density that is constant in time.
Note that thermal equilibrium would not be reached, and some of the ideas
\cite{Sa} that attempt to explain the preponderance of baryons would become
more viable. Some people (including Fred Hoyle \cite{H}) are more
comfortable with this scenario.

Before we turn to the more spectacular aspects of anti-De Sitter physics, it
is worth while to stress, one more time, the theoretical benefits of
negative curvature, however small. The energy spectra of elementary
particles in anti-De Sitter space are positive definite; thus there are
no infrared singularities. Given the pivotal role of the infrared
catastrophy in QCD, I marvel at the fact that no attempt has yet been made
to investigate the effect of negative curvature on the confinement
problem.

\section{Singletons.}

By far, the most dramatic consequence of allowing for a small negative
curvature is the existence of singletons. These are highly degenerate
representations of $SO(3,2)$, with positive energy and thus at first sight
associated with elementary particles. They were discovered by Dirac
in 1963 \cite{Di}. As elementary particles they were at first dismissed
by us on the grounds that these representations have too few states to allow
for the formation of localized wave packets. (One manifestation of this is
the singleton black body spectrum: it turns out to be that of ordinary
particles in 3-dimensional space-time \cite{F75}.) But this property is
precisely what gives singletons their fascinating properties \cite{FF80}.
A free singleton with fixed energy has a well defined angular momentum.
If the energy is large enough to be measurable then the angular momentum
is enormous and the state may be associated with sloshing modes of the
universe; if the angular momentum is small then the energy is of the order
of the curvature and thus too small to be observed.

Our first important observation was that all 2-singleton states are massless;
in fact the action of $SO(3,2)$ on the space of 2-singleton states breaks up
into an infinite direct sum of massless representations.  For example, if
$\varphi(x)$ creates a singleton, then $\varphi(x)\varphi(x)$ creates a
massless particle with spin zero, and $\varphi(x)\partial_\mu\varphi(x)$
creates a photon. This property of singletons, for which there is no analog
in flat space, suggests a model of massless particles as 2-singleton
composites. No interaction and no binding energy is associated with this
type of compositeness; it is just a kinematical fact. Composite
electrodynamics was presented in \cite{FF88} and the linear approximation
to composite gravity in \cite{FF98}.

\section{Singletons and Electroweak Interactions}

Rather than continuing with this review of work that has been published
(references may be found in our last paper with Mosh\'e \cite{FFS99})
I prefer to present some new ideas. Mosh\'e had a very strong belief in the
physical role of singletons. As a tribute to Mosh\'e, Daniel Sternheimer
and I have done our best to vindicate this idea -- feeling that it is
something we owe our friend.

The Standard Model is based on ``the weak group'', $S_W = SU(2)\otimes U(1)$,
and more precisely on the Glashow representation of this group, that
is carried by the triplet ($\nu_e,e_L;e_R$) and by each of the other
generations of leptons. Let us now suppose that
\begin{itemize}
\item[(a)] there are three bosonic singletons
$(R^N R^L;R^R) = (R^A)_{A=N,L,R}$ (three ``Rac''s) that carry the
Glashow representation of $S_W$;
\item[(b)] there are three spinorial singletons
$(D_\varepsilon, D_\mu;D_\tau) = (D_\alpha)_{\alpha = \varepsilon,\mu,\tau}$
(three ``Di''s). They are insensitive to $S_W$ but transform as a Glashow
triplet with respect to another group $S_F$ (the ``flavor group''),
isomorphic to $S_W$;
\item[(c)] the vector mesons of the standard model are Rac-Rac
composites, the leptons are Di-Rac composites, and there is a set
of vector mesons that are Di-Di composites and that play exactly the
same role for $S_F$ as the weak vector bosons do for $S_W$: \\
\begin{eqnarray*}
W_A^B &=&\bar R^BR_A, \\
L_\beta^A &=& R^A D_\beta, \\
F_\beta^\alpha &= &\bar{D}_\beta D^\alpha.
\end{eqnarray*}
\end{itemize}
The vector mesons are associated with conserved currents and fall into a
category of composite particles that was described in Mosh\'e's last paper
\cite{FF98}. There is not any strong evidence, at this time, that $S_F$
is isomorphic to $S_W$, only that $S_F$ has a representation of
dimension 3. The assignments of transformation properties of Di's and Rac's
can be interchanged.

We propose a slightly more economical model; namely we shall identify the
two $U(1)$s with each other. There is only one $U(1)$; the symmetry group
is $SU(2)_W\otimes U(1)\otimes SU(2)_F$. The subgroup $SU(2)_W$ acts on the
Racs, $SU(2)_F$ acts on the Dis, and the hypercharge generator of $U(1)$
acts on both.

Let us concentrate on the leptons ($A=N,L,R$; $\beta=\varepsilon,\mu,\tau$)
\begin{equation}\label{1}
(L_\beta^A) = \left(\begin{array}{ccc}\nu_e & e_L & e_R \\
\nu_\mu & \mu_L & \mu_R \\ \nu_\tau & \tau_L & \tau_R
 \end{array} \right)~.
\end{equation}

It is a remarkable fact that there are very good reasons to believe
that this collection of leptons, precisely three complete generations, is
complete. If leptons are composite, and if lepton fields are related
to bilinears, then the constituents must include both bosons and fermions and
the factorization $L_\beta^A = R^AD_\beta$ is strongly urged upon us by the
nature of the phenomenological summary in Eq.(1).

Fields in the first two columns couple horizontally to make the standard
electroweak current, those in the last two pair off to make Dirac mass-terms.
Particles in the first two rows combine to make the (neutral) flavor
current and couple to the flavor vector mesons.

The Higgs fields have a Yukawa coupling to lepton currents,
\begin{equation}\label{2}
{\cal L}_{\scriptscriptstyle {\mathrm{Yu}}} =
 - g_{\scriptscriptstyle {\mathrm{Yu}}}\bar{L}^\beta_A
L^B_\alpha H_{\beta B}^{\alpha A}.
\end{equation}
The Standard Model was constructed with a single generation in mind,
hence it assumes a single Higgs doublet, and must therefore introduce
three independent Yukawa coupling constants,
$g_{\scriptscriptstyle {\mathrm{Yu}}}H_{\beta B}^{\alpha A} \rightarrow
\delta^\alpha_\beta g_\alpha H_B^A$.

However, an early and remarkable property of weak interaction phenomen\-ology
was electron-muon universality. In a theory based on intermediary
vector mesons, this is expressed as an equality of the coupling constants
of the basic interaction lepton-lepton-meson. In Weinberg-Salam theory,
with its spontaneous breakdown of Yang-Mills symmetry, this equality of
coupling constants is natural and of geometric origin. If, as has been
proposed, the Higgs field also has a geometrical meaning \cite{CL,C93},
then it is natural to suppose that the Yukawa couplings lepton-lepton-Higgs
are also characterized by a universal coupling constant; the same for
electron, muon and (?) tau. This symmetry between leptons is broken only
spontaneously, by the spread of the vacuum expectation values (VeVs)
of the Higgs field.

Symmetry with respect to the group $SU(2)_W \otimes U(1) \otimes SU(2)_F$
does not justify
Eq.(2) with a single Yukawa coupling constant. We shall return to this
point below. It is assumed that
all the neutrinos are lefthanded.

If the great number of components of Higgs fields is unwelcome,
it should be kept in mind that the usual Higgs field is not widely
believed to correspond to an elementary particle. If it is composite,
then there is no reason to expect it to have only two components. In
our model, where leptons and vector mesons are composed of singletons,
it would be natural to suppose that the Higgs field is likewise composed
of singletons, and the topological structure of singleton field theory
even suggests that there may be no elementary particles associated with
the Higgs field in the full dynamical theory of the future.

Nonvanishing vacuum expectation values of the neutral components are
directly related to charged lepton masses,
\begin{equation}\label{3}
g_{\scriptscriptstyle{\mathrm{Yu}}}\langle H_{\alpha R}^{\alpha L}\rangle =
m_\alpha,~ \alpha = \varepsilon,\mu,\tau.
\end{equation}
The coupling constants in $m_\epsilon$ and $m_\mu$ must be equal, but the
third one, associated with the tau, can be
different. If these are the only components with nonvanishing VeV, then the
following
masses are induced for the weak vector mesons:
\begin{equation}\label{4}
m^2(W^\pm) =  g^2 \sum_\alpha \langle H_{\alpha R}^{\alpha L}\rangle^2,
\quad m^2(Z) =
 (g^2+g'^2)\sum_\alpha
\langle H_{\alpha R}^{\alpha L}\rangle^2,\quad
\end{equation}
where $g$ and $g'$ are the two fundamental coupling constants of the
Standard Model ($\tan \theta_W=g'/g$), and
\begin{equation}\label{5}
m^2(C^\pm) = h^2 (\langle H_{\mu R}^{\mu L}\rangle^2 - \langle H_{\epsilon
R}^{\epsilon L}\rangle^2),
\end{equation}
where $h$ is the gauge coupling constant associated with the flavor group
$SU(2)_F$. The theory is still invariant under the three abelian groups
associated with the three lepton numbers,
but only $L_\mu - L_\epsilon$ is gauged, by the flavor gauge boson $C_3$
that remains massless, so far. The hypercharges of the leptons are the
same as in the Standard Model and the hypercharges of the Higgs field
are determined by the postulated invariance of the interaction.

\section{The new developments}

Neutrino oscillations, especially between the two neutrinos associated to
the muon and to the tau, appear to have been firmly established
\cite{SK,BGG,OS,FKM,P}. This suggests non-vanishing neutrino masses but does
not imply it, especially if new flavor changing interactions are not ruled out
\cite{GNPPZ,GPS,JM}.
Nevertheless, there are several attractive mechanisms that account for most
of the data and give masses to at least some of the neutrinos,
without introducing any additional leptons. The most economical assumption
is that the masses of $\nu_\mu$ and $\nu_\tau$ are of the order of .1 eV and
that of $ \nu_e$ even smaller, possibly zero. It was pointed out,
almost 20 years ago \cite{GR,GGN,K}, that there is room for an additional
Higgs field, an $SU(2)$ triplet, and that this implies non-vanishing neutrino
masses. Our proposal is slightly different.

We postulate additional Higgs fields, coupled to leptons in the following way,
\begin{equation}\label{6}
{\cal L}'_{\scriptscriptstyle{\mathrm{Yu}}} =
h_{\scriptscriptstyle {\mathrm{Yu}}} L_\alpha^A L_\beta^B K_{AB}^{\alpha\beta}
+ {\rm h.c.}.
\end{equation}
As in the case of Eq.(3), this formula should not be taken quite literally,
for symmetry does not require that all the
couplings have the same strength.

We now give a nonvanishing vacuum
expectation values to the neutral components (generalizing [GR])
\begin{equation}\label{7}
\langle K_{\scriptscriptstyle{NN}}^{\alpha\beta}\rangle,~ \alpha,\beta =
\epsilon, \mu, \tau.
\end{equation}
This implies a general, symmetric neutrino mass matrix.
In addition, one of the two remaining massless vector mesons becomes
massive. The masses are
$$
m^2(W^\pm)  = g^2(\sum_\alpha\langle H^{\alpha L}_{\alpha R}\rangle^2 + 2
\sum_{\alpha}\langle K^{\alpha
\tau}_{NN}\rangle^2),
$$
$$
m^2(C^\pm)  = 2h^2(\langle H^{\mu L}_{\mu R}\rangle - H^{\epsilon
L}_{\epsilon R}\rangle)^2,
$$$$
m^2 (C^3) = h^2\sum_\alpha\langle K^{\alpha \tau}_{NN}\rangle^2,
$$$$
m^2(Z) = (g^2 + g'^2)(\sum\langle H^{\alpha L}_{\alpha R}\rangle^2 +
4\sum_\alpha \langle K^{\alpha
\tau}_{NN}\rangle^2).
$$

\section{Other predictions of the model.}

The interpretation of the atmospheric neutrino experiments must take into
account, not only neutrino masses and mixing, but also the effect of
the flavor-changing interactions induced by the new vector mesons.
To account for the observed smallness of flavor changing interactions it is
necessary that the new VeV be at least of the order of magnitude of 100 GeV,
so that $h_{{\scriptscriptstyle{\mathrm{Yu}}}}$ must be very
small. Strong constraints on $h_{{\scriptscriptstyle{\mathrm{Yu}}}}$,
and $h$ are imposed by past and future experiments involving
neutral currents. As these become more accurate the model may fail.

\end{document}